\documentclass[letterpaper]{article} 
\usepackage{aaai2027}  
\usepackage[hyphens]{url}  
\usepackage{graphicx} 
\usepackage{natbib}  
\usepackage{caption} 
\usepackage{algorithm}
\usepackage{algorithmic}

\usepackage{newfloat}
\usepackage{listings}
\DeclareCaptionStyle{ruled}{labelfont=normalfont,labelsep=colon,strut=off} 
\floatstyle{ruled}
\newfloat{listing}{tb}{lst}{}
\floatname{listing}{Listing}

\usepackage{booktabs}

\usepackage{amsmath}
\usepackage{cleveref}
\title{The Effect of Perceived Race and Gender on Police Language Use:\\
Experimental Evidence from VR Simulations\\
}
\author{
    Written by AAAI Press Staff\textsuperscript{\rm 1}\thanks{With help from the AAAI Publications Committee.}\\
    AAAI Style Contributions by Peter Patel Schneider,
    Sunil Issar,\\
    J. Scott Penberthy,
    George Ferguson,
    Hans Guesgen,
    Francisco Cruz\equalcontrib\corresponding,
    Marc Pujol-Gonzalez\equalcontrib\corresponding
}
\affiliations{
    \textsuperscript{\rm 1}Association for the Advancement of Artificial Intelligence\\

    1101 Pennsylvania Ave, NW Suite 300\\
    Washington, DC 20004 USA\\
    proceedings-questions@aaai.org
}

\title{The Effect of Perceived Race and Gender on Police Language Use:\\
Experimental Evidence from VR Simulations}
\author {
    Sandra C. Sandoval,
    Navita Goyal\\
    Rashawn Ray,
    Long Doan \\
    Rachel Rudinger,
    Hal Daumé III
}
\affiliations {
    sandracs@umd.edu\\
    University of Maryland
    
}

\begin{document}

\maketitle

\begin{abstract}
Against the backdrop of violence in police interactions with the U.S. public, we explore how deferentially police officers speak to virtual characters depicted as Black adult males in virtual reality (VR) simulations. We evaluate the effect of seeing and communicating with these characters through a causal inference lens, where the assignment of the \emph{Black man} character to a police officer and simulation is the treatment variable. Our (marginal) average treatment effect $ATE$ measures the social impact of the character on the deference of officer statements with each turn of the conversation. Soberingly, we find that most officers speak less deferentially to \emph{Black man} characters, except for White, biracial, and multiracial female officers, especially in settings where the VR character was known to be a suspect. Across a full conversation of a typical VR scene, these marginal $ATE$s can result in notable changes in deference of tone (two to several points difference on a scale of 0-10), above and beyond that due to the initial effect of perceiving a Black male character. Even more disconcerting is that this can contribute to conversation breakdowns that potentially result in violence or danger to both the public and the police. We also explored the capabilities of large language models (LLMs) for ATE estimation. From our methods comparison analysis, including model validation against synthetic data, we provide unique scientific insights on LLM-assisted methodologies for ATE estimation. As such, for ATE estimation with multilevel data with text, we recommend mixed effects models with the inverse propensity treatment weighted (\emph{iptw}) approach, which utilized an LLM for text feature creation. While we also tested LLMs for finetuning prediction models ultimately for ATE estimation, we conclude they are an area for further development and refinement. 

\end{abstract}


\section{Introduction}
How to address the issue of exceptionally high rates of police use of force, police shootings, and fatal police violence in the United States \citep{hirschfield2023exceptionally} is an ongoing and sobering challenge that requires innovative solutions that promote both effective policing as well as the safety of both the police and the public. Such use of force incidents disproportionately affect minority ethnic populations\citep{hirschfield2023exceptionally}. Consequently, in our research we focus on bias in police language use toward Black American men, which is critical as a precursor to aggression \citep{kahn2020social}. Black male adults are proxied by \emph{Black Man} virtual characters in VR settings in our study. 
\begin{enumerate}
\item We first seek to better understand:
\begin{itemize}
\item \emph{RQ1: Do officers talk to Black Man virtual characters differently than they talk to others? If so, how can we best measure the social impact of this?} and 
\item \emph{RQ2: If so, how much of this is influenced by their social beliefs and views on policing (as indicated by their survey responses)?
}
\end{itemize}
\item Second, we explore how large language models (LLMs) can facilitate computing causal effects in data from field experiments with police officers interacting with VR characters.
\end{enumerate}
We evaluate the social impact of seeing a \emph{Black man} character on police language toward these characters in VR lab field experiments. In this setting, 79 police officers from three United States regions (Southern, Mid-Atlantic, and Midwest) interacted with virtual characters, depicting persons of interest or victims, in VR scenes that emulated real life -- a house, a bus stop, and a convenience store. The virtual character's skin tone (\emph{Black} or \textit{White}), gender (\textit{Man} or \textit{Woman}), and the crime scene (VR setting) were randomly assigned for a given officer. The controlled experimental setup included random assignment of the VR character's demographic (ex. \emph{Black Man}) to the police officer's scene simulation.\cite{doan2021evaluation} The demographic of the VR character was our \emph{treatment} variable; this enabled isolation and measurement of the average treatment effect ($ATE$) \cite{sundaram2022importance} on our outcome variable, the deference rating of a set of officer statements. 

$ATE$ is our measure of social impact; a negative $ATE$ indicates less deferential language toward Black males. Knowing the impact of each additional exchange (marginal $ATE$) in a conversation can inform police training on de-escalation in service of dynamically modifying their language over the course of a conversation as needed.  
$ATE$ is challenging to measure in "the wild", such as of police interactions with the public, given common biases in observational data (eg. selection and measurement biases), which risk inaccurate estimates \cite{sundaram2022importance}.

Deference was judged on the transcribed simulations by 5,009 Prolific crowd workers on 7,956 individual officer statements that were made \cite{doan2021evaluation}\footnote{The police statements' deference qualities were judged by annotators. The percent agreement, the degree to which two or more raters provided identical absolute scores \cite{abbott2025achieving} across annotations, were as follows: respect 	0.7666,
formal 	0.7136,
polite 	0.7635,
friendly 	0.7668,
impartial 	0.7496,
deference 	0.7964
}. 
Measurement of the $ATE$s was partially facilitated through the use of AI, specifically the Llama 3.1 8B \cite{Meta_Llama_3.1_8B_2024} large language model (LLM).  
We created the officer statements in our synthetic datasets used to validate our models with the Gemma 4 E 4B \cite{team2024gemma} LLM and scored them  with the RoBERTa \cite{liu2019roberta}, to find the paraphrases that were closest to the original officer statements (see Methodology).

\begin{itemize} 
\item Our contributions include:
\begin{itemize}
\item 
\emph{Our finding that a decrease in deference of language  used by officers may perpetuate and accumulate significantly throughout a conversation. We therefore emphasize that police can and should try to mitigate this adverse social impact by de-escalating tension or conflict in their interactions with the public, including maintaining, recovering or building deference in their language throughout a conversation.  
We call for de-escalation support through policy changes and local public investment in training for police, a form of officer education, which we show is correlated with more deferential police language.}
\item \emph{Scientific insights into 
utilizing large language models (LLMs) in hybrid statistical and AI approaches for the measurement of causal effects over a conversation, with important findings concerning appropriate usage of LLM embeddings and LLM models for statistical analysis of multilevel data with text from VR field experiments with police}
\item \emph{An annotated dataset in which we merged anonymized police survey answers, demographics, and deference ratings data from \citet{doan2021evaluation}, and newly integrated character responses to form full VR simulation dialogues by scene, also including model weights, propensity scores, deference outcome predictions, and other data coding for expanded usability and contextualization.}
\end{itemize}
\end{itemize}
\section{Related Work}\label{sec:related_works}

\label{policing_stragies}
\subsubsection{Language and Social Factors Associated with high Police Use of Force Incidents}\label{global_policing}
Per \citet{kahn2020social}, the majority of White persons hold the "race-crime" implicit stereotype, i.e., associating racial minorities with criminality and violence. Concerningly, this implicit bias influences police decisions to shoot \cite{kahn2020social}. In simulations, participants are more likely to mistakenly shoot unarmed Blacks compared with unarmed Whites \citep{kahn2020social}. In actual traffic stops of Black drivers in the U.S., those ending in arrest, handcuffing, or a search are distinguished by the first 45 words spoken by the officer \cite{rho2023escalated}. Police, as public servants, can prevent these types of escalations. \citet{giles2024policing} note that people feel more rapport with police who are accommodating; whereas, initially disrespectful or overly authoritative comments in police encounters can result in adverse outcomes "...for Black male drivers and, in turn, police–community relations" \citep{giles2024policing}. 
\subsubsection{The Evidence for De-Escalation Training} \label{training_evidence} 
\citet{police2021police} note randomized controlled trials (RCTs) with promising results in support of de-escalation training. A couple of examples include the Integrating Communications, Assessment, and Tactics (ICAT)  program \citep{icat_2025} 
led to reductions in use of force, officer injuries, and civilian injuries, with officers amenable toward the training. The Tact, Tactics, and Trust program \citep{ttt_2026} 
emphasized deliberate, repetitive practice and scenario-based training that officers valued.  Finally, 
\citet{d2021moving} confirmed officer appreciation of de-escalation tactics, such as compromise, safety, and restraint. 
Thus, evidence suggests that police training for de-escalation is critical, practices are trainable and generally welcomed by police. Also, our finding that officers of most demographic backgrounds speak less deferentially to Black male VR characters (or Black men), is validated in the "real world" with evidence from police body camera footage that officers speak less respectfully to Black community members than to White ones \citep{voigt2017language}.

\subsubsection{Language Model Strengths for Predictive Modeling and Growth Areas for Multilevel Data}\label{LLMs_and_causal_est}
LLMs for predictive modeling with text covariates have strengths but key limitations as well. 
In our study, we utilize the LLM, Llama 3.1 8B \cite{Meta_Llama_3.1_8B_2024}, for predictive models in different ways. 
We featurize the text of the conversations between the police officer and the character in a scene. \citet{tang2025understanding} note that this application, using LLM embeddings for downstream tasks, may perform better than feature engineering for high-dimensional regression tasks. 
We also employ the LLM to finetune classification and regression models create model predictions for downstream 
$ATE$ computations. 
\citet{minaee2021deep} argue that deep learning models have surpassed classical machine learning models in text classification. 

Nonetheless, despite widespread adoption of LLMs for statistical analysis via finetuned models and further, the use of downstream LLM-created embeddings, the proper use of LLMs is nuanced for multilevel data; i.e., data with inherent clusters and hierarchical relationships. 
\citet{dickinson2005multilevel} warn against disregarding multilevel data in statistical modeling, and \citet{li2013propensity} show that ignoring multilevel data with propensity score weighting can bias the estimates. Also, \citet{fuentes2022causal} argue for the presence of random slopes in the treatment and the outcome models to properly estimate $ATE$s in multilevel data; in other words, covariate effects should be allowed to vary across clusters (like officers or scenes for us).

LLMs could be stronger at multi-step reasoning \cite{patel2024multi}, in processing and generation related to multilevel data. 
LLMs' optimization for next token prediction may not entail a deeper understanding of data relationships. 
While LLMs perform well with supervised learning tasks such as classification, they have challenges with nuanced unlabeled multilevel data patterns, often taking shortcuts and learning only the easier patterns, and with multi-step planning or reasoning, or "looking ahead". \citep{bachmann2024pitfalls}. We see signs of some of these challenges in our study.




\section{Methodology}\label{sec:methodology}
\begin{table}[!t]
\centering
\scriptsize
\begin{tabular}{p{0.6\columnwidth}c}
\toprule
\textbf{Officer Statements} & \textbf{Deference Ratings} \\
\midrule
ma'am how are you & 9.0 \\
I'm not saying you did just want to talk to you for a minute everything okay & 6.0 \\
is everything okay what are you doing out here today & 6.5 \\
which bus are you waiting for today & 8.0 \\
did you have any problems with anyone today awesome people said there was a verbal argument here are you okay what did you guys argue about & 6.0 \\
\bottomrule
\end{tabular}
\caption{\small Example police officer statements and their crowd worker-assigned deference ratings on a scale of 0-10, from a bus stop scenario.}
\label{tab:officer_statements_and_deference_ratings}
\end{table}
To measure the impact of perceiving a \emph{Black man} VR character on the deference of officer language to the character, we computed the $ATE$ of the character's demographic (race and gender) on the deference ratings of police officer statements. 
Deference is the mean deference score applied to a set of officer statements toward a VR character, an average of the 
ratings for respect, formality, politeness, friendliness, and impartiality (ranging from 0 to 10, as seen in \cref{tab:officer_statements_and_deference_ratings}). See the supplement
for the annotation guidance. The demographic $D=1$ indicated darker skin tone and male gender to be perceived as a \emph{Black Man} VR character; $D=0$ was a catch all for other demographic values of \emph{White Woman, Black Woman, and White Man}, grouped as \emph{All Others}. We estimate the $ATE$ of
\emph{D} on the officer deference of speech, \emph{U} (for officer utterances), in a simulation conversation. We test if the $ATE$ is zero with two-tailed tests at $\alpha = 0.05$.

\subsection{Our Causal Model}
We define our causal model variables as follows:
\begin{itemize}
\small
\item \emph{\textbf{B}: Officer \textbf{b}eliefs (unobserved)
}
\item \emph{\textbf{M}: Officer de\textbf{m}ographics}
\item \emph{\textbf{S}: Answers to \textbf{s}urvey questions about beliefs}
\item \emph{\textbf{U}: Deference $U$ (outcome) of the next officer statements (time step $t + 1$)  to the virtual character}
\item \emph{\textbf{C}: The historical \textbf{c}ontext with dialogue through time $t$ and the average deference of the officer statements (see \cref{tab:officer_statements_and_deference_ratings})}
\item \emph{\textbf{T}: The VR scene/se\textbf{t}ting [e.g. house, bus stop, or convenience store]} 
\item \emph{\textbf{D}: \textbf{D}emographic of the virtual character (treatment variable where $D = 1$ is the \emph{Black Man} perceived demographic)}
\item \emph{\textbf{O}: Officer ID , unique identifier for the officer, implied in the causal graph} 
\end{itemize}
\label{fig:causal_graph_variables_list}

\begin{figure}[!t]
\centering
\resizebox{0.8\columnwidth}{!}{%
\includegraphics[height=3cm]{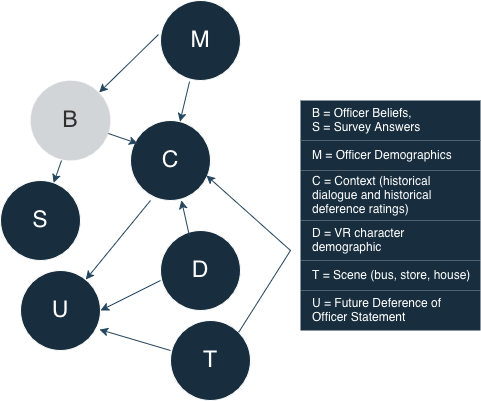}}
\caption{\small \textbf{Causal Graph} Here we depict 
the treatment effect of the VR character on the deference ratings ($U$) of the police officer's next statements in a police scene simulation. 
Confounders also affecting $U$ include $C$, the scene's context including historical dialogue and deference ratings of officer statements, and $T$ for the scene (house, bus stop, or convenience store) of the VR simulation.
Survey variables (S) proxy B, and M represents the officer demographics.
Opaque black circles are observed variables, with the transparent circle ($B$) denoting unobserved officer belief.  
}
\label{fig:Causal_Graph}
\end{figure}
Our causal graph (\cref{fig:Causal_Graph}) illustrates the effect of the treatment $D$ on $U$ (as denoted by the arrow between the two). Since $D$ and $T$ are randomly assigned, they are not caused by other factors. $B$, $C$, $M$, $D$, and $T$ impact what the officer says at any given time. $B$ and $M$ also cause the context, as do $D$ and $T$. 
$S$ approximates $B$ (independent of the rest of the graph).
In computing the $ATE$ for the \emph{Black Man} character demographic ($D = 1$), we sought to understand the treatment impacts on the dynamically changing deference of the officer's language throughout a scene. But, a challenge for the measurement of $D$'s impact on $U$ is the existence of ($B$,$C$,$M$,$T$), confounders between $D$ and $U$. To account for these, we compute $P[D=1 | S, C, M, T]$, the propensity score, or the likelihood of a \emph{Black Man} character. Downstream, these scores are utilized for both methods for $ATE$ estimation we explored. Propensity score weighting of the observations is used in these approaches to create a counterfactual population where covariates are balanced as if under randomization; this then enables us to isolate the effect of $D$ on $U$ through an outcome regression model (or the weighting can be applied after regression in the $ATE$ computation formula). In our outcome model, we estimate the deference of the officer speech $U$ for the next turn in the conversation. The deference predictions from these models serve as inputs to our downstream average treatment effect calculations.

\begin{align}
\small
\color{blue}U_{t+1} &= f \Big(b + \color{violet}\underbrace{d_1D_1}_{\textsf{\tiny Treatment: VR Character Race \& Gender}} \nonumber \\
&\phantom{{} = f \Big(b} + \color{green!50!black}\underbrace{b_2B_2+m_3M_3+s_4S_4+(c_5C_5)_{t}}_{\textsf{\tiny Confounders}} \nonumber \\
&\phantom{{} = f \Big(b} + \color{black}\underbrace{t_2T_2}_{\textsf{\tiny Scene}} \hspace{1.2em} + \color{black}\underbrace{o_2O_2}_{\textsf{\tiny Officer ID}}\Big)
\end{align}
\label{eq: deference_equation} 
We depict the conceptual formula for our deference outcome models in Equation 1 
(applicable to both of our methodological pipelines, I and II).
We regress $U_{t+1}$, the average deference
rating $U$ for the police statements in the next conversational exchange with the character at time $t+1$, 
on the fixed effects in purple and green, and random effects in black. For the propensity models, the formula was almost the same but regressed the binary \emph{D}, the VR character's race and gender, on covariates in the groups shown but excluding \emph{D}.

For both pipelines, we computed the $ATE$ where our observations were for an increasing cumulative window of conversation for a given scene (bus stop, etc.). Thus, each observation contained the  \emph{Context C} variables representing the officer and character's conversation to that point in the dialogue, including the last VR character response and an average (Pipeline I) or list (Pipeline II) of the dialogue's deference ratings for the officer statements to that point, with other covariates as listed in \cref{fig:Causal_Graph}. 


\subsection{Note on Comparison of Statistical Methods for ATE Computation}
We tested two primary causal inference computation methods for deriving the $ATE$ of a perceived \emph{Black man} character on officer deference of language to the character. The first was inverse propensity treatment weighting (\emph{iptw}) \cite{rosenbaum1983central} and the second is termed a "doubly robust" approach to $ATE$ estimation \cite{funk2011doubly}. While we define both approaches in the next section for Pipeline I, we compared both approaches for each pipeline.

\subsection{Pipeline I: Mixed Effects Models with LLM-Featurized Dialogue (and ATE Computation Methods for both Pipelines)}\label{sec:pipelineI}
For Pipeline I, to ensure statistically robust estimation of the average treatment effect, we account for multilevel data, such as repeated measures and non-independence of utterances per officer in a scene, or correlation of scene observations over time, through the use of mixed effects regression models\footnote{We utilized the Generalized Linear Mixed Effect Model (glmer) function of the R lme4 package version 1.1.38 for logistic regression to estimate propensity scores which we utilized for \emph{iptw} in Pipeline I. We utilized the Linear Mixed Effects Models (lme) function of the nlme package version 3.1.168 in R to model our deference outcome variable with regression.} \cite{gelman2007data}. 

\textbf{Inverse Propensity Treatment Weighting ATE}
We performed an \emph{iptw} approach to assess the $ATE$ of the demographic of the character on the deference of police officer statements. 
To do so, we first gathered propensity scores ($\pi$), or the probabilities of the treatment for the observation at each time step in the dialogue, via a mixed effects logistic regression model \citep{bates2015fitting}. We then fit a mixed effects linear regression model on our outcome, average deference of the next time period $U_{t+1}$, with the same covariates plus an additional one for the demographic treatment \emph{D} (see formula 1). 
This \emph{iptw} approach utilized the previously obtained propensity scores (see equations 2 and 3 below, where $D=1$ for the \emph{Black man} character, and $D=0$ for \emph{All Others}). 
We trimmed the dataset to exclude extreme propensity weights resulting in scores for the field experiments data ranging from 0.01 to 0.99 inclusive (removing observations highly likely or unlikely to receive the treatment). 

After refitting, with the outcome predictions from our mixed effects deference regression model, we computed the $ATE$ of $D=1$ on the average deference at time period $t+1$ by taking the difference between the mean of the treated bootstrapped resamples and that of the control group, as follows\footnote{
After obtaining the inverse propensity weighted \emph{iptw} deference model predictions, we performed bootstrap resampling (with replacement) for 1000 iterations. 
We calculate statistical significance and report Benjamini Hochberg corrected ATEs for 11 total tests.}:
\begin{align}
\small
y_{\textit{treated}(D=1)} &= \frac 1 {\hat\pi} (y) \\y_{\textit{control}(D=0)} &= \frac 1 {1 - \hat\pi} (y)
\end{align}
\label{eq:iptw equation}
{
\begin{equation}
    ATE_{\emph{iptw}} = \bar{\hat{y}}_{\textit{treated}} - \bar{\hat{y}}_{\textit{control}}
\end{equation}
\label{ATE_formula}
}

The Pipeline I approach was hybrid, utilizing a statistical mixed effects model while pulling in the  \emph{Context C} PCA-reduced dialogue features as covariates \footnote{PCA with scikit-learn 1.7.2.for use with R software.}; we used PCA to reduce the embeddings output from the last token of each input sequence, which represented each entire prompt, from the final layer of the the Llama 3.1 8B transformer model.
We additionally perform the deference outcome model regression without propensity score weighting as shown in \emph{Table 3}, 
to assess the correlations of our other covariates with the deference ratings for our second research question.

\textbf{Doubly Robust ATE}
For comparison purposes, we additionally computed
\cite{funk2011doubly}'s  doubly robust $ATE$ estimation equation (\cref{eq:doubly_robust_equation}) with the logistic and deference prediction outputs, as seen below, in three steps. The doubly robust method is intended to mitigate potential biases in $ATE$ estimation when one of the models may be misspecified, such as from un-(or only partially-) measured confounding variables influencing the officer deference. 

In the doubly robust formula, we term the propensity weights $\pi$, the true average deference ratings $y$, the predicted (unweighted) deference ratings $\hat{y}$ and $d$ for the binary labels for character demographic $D$ \citep{funk2011doubly}:

\begin{align}
\small
\Psi_{\textit{treated}} &= \frac 1 {\hat\pi} \Big(yd - \hat{y}_1(d - \hat\pi)\Big)  \\
\Psi_{\textit{control}} &= \frac 1 {1 - \hat\pi} \Big(y (1 - d) + \hat{y}_0 (d-\hat\pi)\Big) \\
ATE_{dr} &= \bar{\Psi}_{\textit{treated}} - \bar{\Psi}_{\textit{control}}
\label{eq:doubly_robust_equation}
\end{align}

\subsection{Pipeline II: LLM-Finetuned Prediction Models} \label{sec:pipelineII}
In the case of Pipeline II, 
the \emph{iptw} and doubly robust $ATE$ estimations were computed 
and the steps were similar, but we finetuned separate (unweighted) deference and propensity models utilizing the Llama 3.1 8B base model \cite{Meta_Llama_3.1_8B_2024}, then utilized each model's predictions downstream as inputs to both the \emph{iptw} model (synthetic data only, given less desirable model quality) and doubly robust $ATE$ estimation models for comparison and model validation. For our finetuned propensity and deference outcomes, we attempted to address the multilevel nature of our data utilizing custom regression heads for the architecture such that the standard single linear layer was replaced with a feedforward network \cite{vaswani2017attention} consisting of multiple layers (specifically a multilayer perceptron), including activation functions for nonlinear data relationships. Specifically, for the finetuned propensity model (classification model) regressing D on the causal model covariates, we utilized an MLP with three linear layers with Gaussian Error Linear Unit (GeLU) nonlinear activation layers in between, minimizing a cross entropy loss function. For the finetuned deference model, we utilized an MLP with two linear layers and a ReLU (Rectified Linear Unit) nonlinear activation layer, minimizing a mean squared error (MSE) loss; this was an unweighted loss in the case of the basic deference predictions required for the doubly robust $ATE$ estimation, and an inverse propensity score-weighted loss in the case of the \emph{iptw} $ATE$ estimation.

 
For this approach, our observations were set up with a temporally increasing window of dialogue as before, but in the form of prompts (see the supplement for an example prompt). 
The covariates included in the text prompt were those described for Pipeline I, and also included a list of the scene dialogue's historical deference ratings. 
\subsection{Model Validation Approach}
Since we had no known ground truth treatment effect to validate that our models would identify such an effect if there was one, we created partially synthetic datasets to test and compare the effectiveness of our models with an artificial treatment effect. To do this, we created synthetic text mirroring the police officer statements from the field experiments. We recreated the original police statements and new versions of the cumulative conversations which utilized LLM-synthesized paraphrases of the original officer statements for the police portions. Our synthetic data was based only on the \emph{Black man character} simulation conversations. This data was further split into two classes such that police officer statements in class 0 were  artificially made more deferential, whereas those in class 1 were intended to mirror the original level of deference by the officer (this paralleled class 1 for the original \emph{Black man} character treatment and class 0 for the \emph{All Others} group). 

We synthesized the paraphrases through prompting and in-context learning. The latter included providing the LLM example police statements that demonstrated different deference levels, with Gemma 4 E 4B \cite{team2024gemma} and scored them with RoBERTa \cite{liu2019roberta}-generated deference scores after finetuning RoBERTA with the original deference data. The winning paraphrase for an officer statement was the one closest in scored deference to the original deference level for the class 1 statements, or closest to the perturbed target (increased) deference score for the class 0 statements. The perturbed target deference value was computed by taking the true deference score of the original police officer statement and adding a delta ($\Delta$) to increase the deference value. The delta was generated by sampling from a geometric distribution such that some values were perturbed a small amount (or not at all) and some values larger to potentially reflect a more realistic range of deference scores. We created three synthetic datasets, to test different levels of increased deference, sampling the geometric distribution for the deltas with ${p=0.2, p = 0.3, p= 1}$ where the greatest delta was expected from $p=0.2$ and no delta from $p=1$. 


\section{Results}\label{sec:results}

We found our Pipeline I mixed effects \emph{iptw} model results to be most reliable for answering our research questions. This was based on a rigorous testing of our pipelines and methods, in which we evaluated their abilities to detect treatment effects as well as their model quality results, across our field experiments and synthetic data sets. 
\setlength{\tabcolsep}{4pt}
\begin{table}[t]
\small
\centering
\label{tab:pipelines}
\begin{tabular}{lcccc}
\toprule
& \multicolumn{2}{c}{Pipeline I} & \multicolumn{2}{c}{Pipeline II} \\
\cmidrule(lr){2-3} \cmidrule(lr){4-5}
Dataset & \emph{iptw} & \emph{dr} & \emph{iptw} & \emph{dr} \\
\midrule
Field experiments     & Y & N & N & N \\
Synthetic (geom. $p=0.3$)   & Y & N & N & Y \\
Synthetic (geom. $p=0.2$)   & Y & N & N & Y \\
Synthetic (geom. $p=1$)$^*$     & Y & N & N & Y \\
\bottomrule
\end{tabular}
\caption{\small\textbf{Significance for Two Pipelines} This table indicates the presence of statistically significant $ATE$s by dataset tested and Pipeline method. We saw more subgroups (ex. officer demographic) with statistically significant \emph{iptw} results, but some statistically significant doubly robust results for the Pipeline II finetuned models as well (mostly for geometric $p=2$ derived perturbations). 
$^*$Note for the geometric $p=1$ derived synthetic data, the \emph{dr} Pipeline II $ATE$s were improbable (outside 0-1).}
\end{table} 
\label{signif_for_2_pipelines}

\subsection{Results from  Pipeline I, Mixed Effects Models with LLM-Featurized Text}
\subsubsection{The Perceived \emph{Black Man} as Virtual Character appears to Causally Influence Police Officer Deference of Speech}

\begin{table}[!t]
\centering
\small
\begin{tabular}{llrrr}
\hline
 Subgroup                                  &    N &   $ATE$ &     p &   BH\\
\hline
 officer\_black\_female  &  141 &      -0.258 & 0.000 &      0.027          \\
 officer\_black\_male                               &  815 &      -0.197 & 0.000 &      0.009 \\
 officer\_bimulti\_race\_female   &  107 &       0.188 & 0.041 &      0.045  \\
 officer\_bimulti\_race\_male      &   70 &      -0.180 & 0.009 &      0.041 \\
 officer\_latina\_female        &   53 &      -0.200 & 0.002 &      0.036  \\
 officer\_white\_female         &   52 &       0.414 & 0.000 &      0.018  \\
officer\_white\_male                                & 1033 &      -0.034 & 0.172 &      0.050  \\
 overall                                       & 2271 &      -0.107 & 0.000 &      0.014 \\
 bus\_scene                                     &  725 &      -0.248 & 0.000 &      0.004 \\
 house\_scene                                  &  786 &      -0.159 & 0.000 &      0.023  \\
 store\_scene                                   &  760 &       0.105 & 0.000 &      0.032  \\
\hline
\end{tabular}
\caption{\small\textbf{Pipeline I (mixed effects \emph{iptw} on VR Field Experiments data)} This table reports the average treatment effect ($ATE$) (scale: 0-10) of the $D = 1$ (\textit{Black Man}) on the deference ($U_{t+1}$) of the police officer statements overall and by subgroup, utilizing the \emph{iptw} approach for $ATE$ calculation reported with Benjamini-Hochberg (BH) corrections for multiple testing. The BH corrections critical values were compared to (and needed to be larger than) the p values to be statistically significant, for 11 total subgroup tests. 
The "overall" value for \emph{Subgroup} includes all observations for all possible subgroups.}
\label{table:ATE_mixedeffects_utterance}
\end{table}

\begin{table}[!t]
\centering
\small
\begin{tabular}{lrr}
\hline
Covariates & Coeff. & p \\
\hline
(Intercept) & 4.231 & 0.000 \\
deference\_historical\_avg & 0.244 & 0.000 \\
sees\_blacks\_as\_hard\_working & -0.019 & 0.670 \\
how\_much\_black\_discrim & -0.065 & 0.019 \\
sees\_blacks\_as\_intelligent & 0.041 & 0.279 \\
has\_family\_who\_are\_police & 0.138 & 0.118 \\
officer\_female & 0.198 & 0.088 \\
officer\_bi\_multi\_racial & 0.015 & 0.926 \\
officer\_latino & -0.153 & 0.511 \\
officer\_anglo & 0.074 & 0.406 \\
officer\_has\_BA & 0.340 & 0.033 \\
officer\_has\_MA & 0.535 & 0.014 \\
my\_trust\_in\_others & -0.008 & 0.796 \\
my\_police\_identity & -0.006 & 0.886 \\
my\_racial\_group\_views & 0.065 & 0.123 \\
\hline
\end{tabular}
\caption{\small\textbf{Pipeline I mixed effects unweighted regression on officer deference (on VR field experiments data)} Distinct from our causal effect analysis in \cref{table:ATE_mixedeffects_utterance}, here we reference a subset of correlation results to investigate RQ2. This unweighted regression is on the deference rating of the officer's next statement, controlling for the previous average deference rating and the cumulative conversation text for the scene, among other variables. 
Some covariates are not shown, including the 50 principal components 
resulting from PCA on the LLM-featurized text. 
\textbf{N=3897} observations at the utterance level, i.e. an officer/character dialogue exchange. Number of groups: \textbf{70 officers} and \textbf{208 unique simulated scenes} by officer resulting from three basic scene types. 
}
\label{table:ATE_mixedeffects_reg_output}
\end{table}

In reference to RQ1, we find that police officers do speak differently to \emph{Black Man} virtual characters versus all other characters (\emph{All Others}, including White and Black females, and White males, our composite control group) in VR simulations of real-life police scenarios. 

In \cref{table:ATE_mixedeffects_utterance}, we show the results (scale of 0 to 10) from the \emph{iptw} mixed effects deference prediction model. 
Here we see, in the overall (across all N observations), store and bus scene subgroups, that the
$ATE$s 
are negative and statistically different than zero. 
While these appear like small effects, this is because they are marginal effects that represent the average change in deference for each additional exchange between officer and character, above and beyond the initial effect from seeing a \emph{Black Man} VR character. In aggregate, they can be substantial. For example, for a scene having 16 officer to character exchanges, 
an $ATE = -.248$ for the \emph{bus} scene suggests a decrease in deference of almost 4 points from the beginning to the end of the conversation. 
The positive $ATE=0.105$ for the store scene is reasonable as officers knew the character in that scene was a victim. 

With respect to officer demographics (race and gender), we see negative impacts (though smaller for male) for Black officers, at $ATE = -0.366$ for Black female officers and $ATE = -0.165$ for Black male officers. 
Notably, only the female officer demographics (white female officer $ATE = +0.414$, biracial and multiracial female officer $ATE=0.188$) appear to have more deferential tones to a \emph{Black Man} character; 
again, in aggregate for the scene these can result in several points increase in deference. 

Finally, our same model results against our synthetic data ($n=414$, geometric $p=0.3$), where we inserted an artificial treatment effect, similarly show negative effects of the perceived \emph{Black man} character (class 1) across all scenes for the \emph{overall} group, the \emph{house scene}, the \emph{store} scene, and the \emph{bus stop} scene (respectively, on a scale of 0 to 1, these were $ATE=-0.090$, $ATE=-0.078$,$ATE=-0.117$, and $ATE=-0.052$; note that in contrast, the ATEs for the field experiments are reported on a scale of 0 - 10). Additionally, the perceived Black man character manifests in an $ATE-0.090$ and an $ATE =-0.089$ for Black male officer and White male officers, respectively. These results against the synthetic data demonstrate that the Pipeline I mixed effects \emph{iptw} model with LLM-featurized text applied to the field experiments data is valid. When there is a ground truth treatment effect in the data, the model is able to detect it.


\subsubsection{Officer Background Explains more Deferential Statements to VR Characters in Police Scenario Simulations}\label{sec:rq2_correlations}
Regarding RQ2, in which we wanted to know how much deference is associated with officer beliefs or attitudes, we find 
that the association of officers' social beliefs or attitudes (reported in their surveys) with officer deference of language is not statistically significant in our models. 
However, having at least a Bachelor of Arts (BA) or a Master of Arts (MA) is associated with an increase of 0.34 and 0.54 points (scale of 0 - 10) in police deference ratings , respectively. It is possible that the police attitudes and beliefs reflected in the survey do not play a significant role in their deference of tone due to subjective police self-reporting, among other reasons. 
Our finding of a significant relationship between education levels and the officer's choice of future wording highlights the important role that learned behaviors, such as language choice, can play in communications and rapport. De-escalation training, something police are amenable to as per our Related Work, can be viewed as a form of education that could be similarly correlated with deferential language choice.
\subsection{Results from  Pipeline II, Fine-Tuned Models with \emph{doubly robust} ATE estimation}
For the field data, the DR model for Pipeline II yielded no statistically significant results; this appears partially due to the more subtle treatment effects in the field  data, and which were an order of magnitude bigger in the synthetic data sets (field $ATE$s: scale 0-10,synthesized $ATE$s: scale 0-1). For the synthetic data (see Table 5), there should have been a negative ATE overall, across scenes and for two officer demographics (White male and Black male), but only the negative White male $ATE$ was detected (geometric $p = 0.3$). The other known effects were only detected with the higher treatment effect (geometric $p = 0.2$). For the synthetic data (geometric $p=0.3$) with a smaller expected delta or increase in deference, we found only a single statistically significant effect of the perceived \emph{Black man} character on police deference of language. 
The Table 5 results suggest that Pipeline II may be suitable for detecting larger treatment effects in the case of the the doubly robust approach; however, model quality remains important.\footnote{Table 5 absolute $ATE$ values are lower than the actual increased deference deltas given the data processing pipeline, where the predicted next deference score is a function of the context, among other variables. We also allow for 0 deference perturbations in our increased deference class to make the conversations realistic; these constituted a significant minority of the deltas.} See full results in the supplement. 
\begin{table}
\small
\begin{tabular}{lcc|cc|cc|c}
\toprule
& \multicolumn{2}{c|}{Actual Def.$\uparrow$}
& \multicolumn{2}{c|}{$ATE$ Pipe. I}
& \multicolumn{2}{c|}{$ATE$ Pipe. II}
& \\
& &
& \multicolumn{2}{c|}{(ME + \emph{iptw})}
& \multicolumn{2}{c|}{(FT + \emph{dr})}
& \\
\cmidrule(r){2-3}
\cmidrule(r){4-5}
\cmidrule(r){6-7}
\cmidrule(l){8-8}
G.\ $p$
& Overall
& $\Delta>0$
& WM
& BM
& WM
& BM
& $p$ \\
\midrule
0.3 & 0.15 & 0.19 & -0.09 & -0.09 & -0.07 & N.S. & $<.001$ \\
0.2 & 0.18 & 0.20 & -0.13 & -0.10 & -0.40 & -0.09 & $<.01$ \\
1.0 & 0.07 & N/A & -0.03 & N.S. & OOS & -0.24 & $<.01$ \\
\bottomrule
\end{tabular}
\caption{\small\textbf{Model Validation on Synthetic Data} (Scale: 0 - 1) On the left, we show the average actual increases in deference ($\Delta$) or perturbations resulting from sampling the geometric distribution (G.p) with the p values shown. All $\Delta$ estimates had standard errors ranging from $\pm0.0065$ to $\pm0.0067$. Geometric $p=1$ corresponds to no intended increase in deference, and $p=0.2$ generated the greatest intended increase. We also show $ATE$ on the officer statement deference by Pipeline I (Mixed Effects-ME with \emph{iptw}), Pipeline II (Fine-Tuning-FT and Doubly Robust-DF estimation), and police officer demographic.\emph{BM} denotes Black male officer and \emph{WM} denotes White male officer. Final column p indicates level of significance for ATEs; not significant is indicated by N.S. $ATE$ estimates outside the 0-1 scale have entries of OOS.}
\end{table}

\subsection{Model Quality Comparisons between Pipeline I and Pipeline II}
For the field experiments data, comparing the model quality of the finetuned deference outcome prediction model (Pipeline II) to the mixed effects deference outcome model (Pipeline I) quality in Table 6, the latter appears more reliable than the finetuned model with respect to the proportion of variance explained (coefficient of determination $R^2= 0.15$ for the mixed effects model versus $R^2 = 0.02$ for the LLM-finetuned model, almost ten times higher. This contrast is similar for the Pipeline I and Pipeline II for the synthetic datasets. This suggests that the finetuned model's predicted outcomes may be influenced by more noise or other unmeasured variables than the mixed effects one, or that the LLM-based regression model, in spite of the mentioned architecture changes, is not sufficiently capturing possibly non-linear relationships. Pearson's correlation coefficient for the mixed effects model is very high ($r = 0.39$) for the mixed effects model versus for the finetuned model ($r = 0.20$). Similarly, Pearsons's $r$ for the Pipeline I \emph{iptw} models are much higher than for Pipeline II. In contrast, other metrics shown are comparable for the two pipelines.
This is plausible given the supervised deep learning model strengths outlined in the Related Work. The \emph{iptw} results for Pipeline II on synthetic data had ( $R^2 < 0$), thus were not implemented on field data.

\begin{table}[t]
\centering
\small
\setlength{\tabcolsep}{3pt}
\label{tab:regression-quality}
\begin{tabular}{lrrrrrr}
\toprule
Model & $N$ & MAE & RMSE & $R^2$ & $r$ & $p$ \\
\midrule
I Field (\emph{iptw})    & 2271 & 0.85 & 1.12 & 0.15 & 0.39 & $<.001$ \\
II Field (reg)    & 1634 & 0.87 & 1.13 & 0.02 & 0.20 & $<.001$ \\
I Synth (\emph{iptw})    & 414  & 0.07 & 0.08 & 0.40 & 0.63 & $<.001$ \\
II Synth (reg)    & 196  & 0.08 & 0.10 & 0.13 & 0.38 & $<.001$ \\
II Synth (\emph{iptw})   & 83   & 0.17 & 0.20 & -2.63 & -0.01 & .94 \\
\bottomrule
\end{tabular}
\caption{\small\textbf{Deference Outcome (Regression) Model Quality Metrics}. This table contrasts model performance on the deference score prediction task, with results from the Pipeline I mixed effects model with \emph{iptw} (labeled I) and the Pipeline II Fine-Tuned models (II), both the regular non-weighted regression (reg) and \emph{iptw} approaches (synthetic only for Pipeline II \emph{iptw}) for both field experiments and sythetic data. Synthetic data metrics are shown only for the $p=0.3$ perturbation set. Metrics shown include Mean Absolute Error ($MAE$), Mean Squared Error ($MSE$), Root Mean Squared Error ($RMSE$), and coefficient of determination ($R^2$), Pearson's $r$ correlation coefficient, $p$ value for Pearson's $r$. N for Pipeline II is for the test splits.}
\end{table}
\label{tab: def_model_qual_metrics}

\section{Conclusion} \label{sec:conclusion}
In this study, with data from VR field experiments with U.S. police officers, we evaluated the social impact of perceiving a \emph{Black man} character on officer language in VR policing scenarios, quantified through $ATE$ estimation of the change in the deference of the officer's next exchange with the character.
Concerningly, we show that police across most demographic groups speak less deferentially with each consecutive statement (average marginal $ATE$) to a  \emph{Black Man} character, than to others. A grave implication of this is that the accumulation of these marginal decreases in deference can result in conversation breakdowns that risk preceding dangerous conflict.   
In contrast, for White, bi and multiracial female officers we showed an increase in the officer's language deference. 
These findings were possible with our Pipeline I, mixed effects \emph{iptw} models with LLM-featurization of the simulation conversations, shown to be valid with synthetic data. 

We highlight scientific insights for challenges related to the use of LLMs for modeling to facilitate $ATE$ computations for multilevel field data, which we experienced even with our use of custom regression heads with multilayer perceptrons (MLPs), commonly used in NLP to account for non-linear relationships in data, multilevel with text in our case. In addition to our $ATE$ findings, our model quality metrics lead us to recommend our Pipeline I (mixed effects with \emph{iptw} and LLM text featurization), over the Pipeline II approach (fine-tuned prediction models with all data and text passed as prompts). Pipeline II is potentially reliable for large synthetic treatment effect sizes (not found in the field data), but inconsistent.  
These methodological findings suggest caution with the use of LLM fine-tuned prediction models for ATE estimation applied to multilevel data with text, particularly from real-world field experiments with humans or observational data.

With the insights from this study, we make the case for local policy changes or human capital investments such as for officer training in de-escalation via police scene simulations, which hold promise as discussed in our \emph{Related Work} section. We call for police training that encourages officers to build, maintain or recover deference in the course of an interaction with public citizens. 




\appendix


\bibliography{aaai2027}


\end{document}